\documentclass[12pt]{article}
\usepackage{amssymb,amsmath}
\usepackage[noblocks]{authblk}
\usepackage[top=0.75in, bottom=0.75in, left=0.75in, right=0.75in, dvips]{geometry}
\usepackage{caption}
\usepackage{graphicx}
\usepackage{epstopdf}
\begin{document}
\textwidth 10.0in
\textheight 9.0in
\topmargin -0.60in
\title{The Effective Potential in Massless Theories}
\author[1,2]{D.G.C. McKeon}
\affil[1] {Department of Applied Mathematics, The
University of Western Ontario, London, ON N6A 5B7, Canada}
\affil[2] {Department of Mathematics and
Computer Science, Algoma University,\newline Sault St.Marie, ON P6A
2G4, Canada}

\maketitle                                 
   
\maketitle
\noindent
email: dgmckeo2@uwo.ca

\begin{abstract}
The effective potential $V$ is a massless self-coupled scalar theory and massless scalar electrodynamics is considered.  Both the $\overline{MS}$ and Coleman-Weinberg renormalization schemes are examined.  The renormalization scheme dependence of $V$ is determined.  Upon summing all of the logarithmic contributions to $V$, it is shown that the implicit and explicit dependence on the renormalization scale $\mu$ cancels.  In addition, if there is spontaneous symmetry breaking, then the dependence on the background field $\Phi$ cancels, leaving $V$ flat but with non-perturbative contributions.  The quartic scalar coupling in the Coleman-Weinberg renormalization scheme consequently vanishes.
\end{abstract}

\section{Introduction}

The non-trivial ground state of the scalar Higgs field is responsible for the mass of the weak vector Bosons as well as the Fermions (though the Stueckelberg mechanism for mass generation could in principle be operative with any $U(1)$ vector Boson [1, 2]).  This was first noticed at the classical level [3-5], but the possibility of the ground state also being affected by quantum effects was later considered [6-9].

There being inherent ambiguities in any perturbative calculation of loop effects in quantum field theory, one is led to the renormalization group (RG) equations [10-12]. These lead to the possibility of summing those parts of higher loop effects involving logarithms of the renormalization mass scale $\mu$ [13-14].  Indeed, it has also proved possible to sum all of these logarithmic contributions to the effective potential $V$ so that $V$ is determined by the log independent contributions and the $RG$ functions.  When this summation is combined with the condition that $V$ has a minimum at some non-vanishing value $v$ of the scalar field $\phi$, it has been shown that $V$ in fact must be independent of $\phi$; this occurs in a simple self-interacting scalar model, scalar electrodynamics is [15, 16], a massive self-interacting model [17] and a massive model which involves interactions between the scalar and other fields [18].  This result is consistent with the general result that $V$ must be convex [19-21], a condition not satisfied by the classical ``Mexican hat'' potential.

It has also been shown that when computing loop contributions to a variety of processes [22-24], the summation of all logarithmic contributions by use of the RG equation leads to full cancellation of $\mu$ dependence between the implicit and explicit dependence on $\mu$.  In addition, the RG equations that follow from ambiguities arising when one uses a mass independent renormalization scheme $(RS)$ make it possible to find a $RS$ in which either the loop effects are absorbed into the RG functions, or the RG functions themselves only receive a finite number of contributions.

In this paper, we first use the RG equation to sum all logarithmic corrections to $V$ when there is only a massless scalar field with a quartic coupling.  This leaves us with $V$ being expressed in terms of the log independent contributions and free of any dependence on the renormalization scale $\mu$.  The RS dependence of $V$ is then considered, so that $V$ can be expressed in terms of the coefficients of the RG function (which characterize the RS [25, 26]) and a set of RS invariants.  Upon requiring that $V$ be at an extremum when the scalar field has a value $v$, we find that if $v \neq 0$ then $V$ is independent of the scalar field (ie, it is ``flat'').  This is consistent with the theorem that $V$ must be convex [19-21].  A flat potential implies that the renormalized quartic coupling, when it is defined using Coleman-Weinberg (CW) renormalization [6], vanishes.  We also find that $V$ contains non-perturbative contributions.

A similar analysis is applied to massless scalar electrodynamics.

\section{Renormalization Group Summation in the Massless Scalar Model}

If one uses the $\overline{MS}$ RS in a model with the classical action
\begin{equation}
\mathcal{L} = \frac{1}{2} \left(\partial_\mu \phi\right)^2 - \frac{\lambda}{4!} \phi^4
\end{equation}
then the effective potential has the form
\begin{equation}
V \left( \lambda , \phi , \mu\right) = \sum_{n=0}^\infty \sum_{m=0}^n T_{n,m} \lambda^{n+1} L^m \phi^4 
\end{equation}
where $L = \ln(\lambda\phi^2/\mu^2)$ with $\phi$ being a constant background field and $\mu$ being the renormalization mass scale.  Since $\mu$ is unphysical, $V$ satisfies the RG equation
\begin{equation}
\left( \mu^2 \frac{\partial}{\partial \mu^2} + \beta (\lambda) \frac{\partial}{\partial\lambda} + \phi^2 \gamma(\lambda) \frac{\partial}{\partial\phi^2}\right) V = 0
\end{equation}
where
\begin{subequations}
\begin{align}
\beta (\lambda) &=  \mu^2\frac{d\lambda}{d\mu^2} = -b\lambda^2 \left(1 + c\lambda + c_2 \lambda^2 + \ldots\right) \\
\gamma (\lambda) &=  \frac{\mu^2}{\phi^2}\frac{d\phi^2}{d\mu^2} = f\lambda\left(1 + g_1 \lambda + \ldots\right).
\end{align}
\end{subequations}
If we write eq. (2) in the form
\begin{equation}
V = \sum_{n=0}^\infty A_n (\lambda) L^n\phi^4
\end{equation}
where
\begin{equation}
A_n (\lambda) = \sum_{k=0}^\infty T_{n+k,n}\lambda^{n+k+1}
\end{equation}
then eq. (3) implies that
\begin{equation}
A_n(\lambda) = \frac{1}{n} \left[ \hat{\beta} (\lambda) \frac{\partial}{\partial\lambda} + 2 \hat{\gamma}(\lambda)\right] A_{n-1}(\lambda)
\end{equation}
where
\begin{subequations}
\begin{align}
\hat{\beta} &= \frac{\beta}{1 - \beta/\lambda - \gamma}\\
\hat{\gamma} &= \frac{\gamma}{1 - \beta/\lambda - \gamma}.
\end{align}
\end{subequations}
If now
\begin{equation}
A_n(\lambda) \equiv \exp - 2 \left[ \int_0^\lambda dx \frac{\gamma (x)}{\beta(x)} + \int_0^\infty dx \frac{fx}{bx^2(1+cx)}\right] B_n(\lambda)
\end{equation}
then by eq. (7)
\begin{equation}
B_n (\lambda) = \frac{1}{n} \hat{\beta}(\lambda) \frac{\partial}{\partial\lambda} B_{n-1} (\lambda).
\end{equation}
(The second integral in eq. (9) is an infinite constant designed to ensure that the argument of the exponential is finite [25].)
If $\eta$ satisfies
\begin{equation}
\frac{d\lambda}{d\eta} = \hat{\beta}(\lambda)
\end{equation}
then eq. (10) becomes
\begin{equation}
B_n (\lambda(\eta)) = \frac{1}{n}\frac{d}{d\eta} B_{n-1} (\lambda(\eta))
\end{equation}
which upon iteration leads to
\begin{equation}
B_n (\lambda(\eta)) = \frac{1}{n!} \frac{d^n}{d\eta^n} B_0 (\lambda(\eta)).
\end{equation}
We thus see that together eqs. (5, 9, 13) lead to
\begin{align}
V &= \sum_{n=0}^\infty \phi^4 \exp -2 \left[ \int_0^\lambda dx \frac{\gamma(x)}{\beta(x)} + \int_0^\infty dx \frac{fx}{bx^2(1+cx)}\right]
 \frac{L^n}{n!} \frac{d^n}{d\eta^n} B_0 (\lambda(\eta))\nonumber \\
& = B_0(\lambda (\eta + L)) \phi^4 \exp -2 
\left[ \int_0^\lambda dx \frac{\gamma(x)}{\beta(x)} + \int_0^\infty dx \frac{fx}{bx^2(1+cx)}\right].
\end{align}
Eqs. (8a, 11) show that
\begin{equation}
\eta + L = \int_0^\lambda dx \frac{1-\beta(x)/x - \gamma(x)}{\beta(x)} + \ln
\left( \frac{\lambda \phi^2}{\mu^2}\right) + K
\end{equation}
where $K$ is a constant of integration chosen so that $\eta$ is finite.

We also know from eqs. (4a,b) that [25]
\begin{equation}
\ln \left( \frac{\mu^2}{\Lambda^2} \right) = \int_0^{\lambda\left(\ln \frac{\mu^2}{\Lambda^2}\right)} dx \frac{1}{\beta(x)} + \int_0^\infty dx \frac{1}{bx^2(1+cx)}
\end{equation}
and as
\begin{equation}
\frac{1}{\phi^2}\frac{d\phi^2}{d\lambda} = \frac{\gamma(\lambda)}{\beta(\lambda)}
\end{equation}
we also have
\begin{equation}
\phi^2 = \Phi^2 \exp \left[ \int_0^{\lambda\left(\ln \frac{\mu^2}{\Lambda^2}\right)} dx \frac{\gamma(x)}{\beta(x)} + \int_0^\infty dx \frac{fx}{bx^2(1+cx)}\right].
\end{equation}
In eqs. (16, 18) $\Lambda^2$ and $\Phi^2$ are constants that arise in the course of integrating eqs. (4a, 17).

Together, eqs. (16, 18) reduce eq. (15) to
\begin{align}
\eta + L &= \ln \left( \frac{\mu^2}{\Lambda^2}\right) + \ln \left( \frac{\Phi^2}{\mu^2}\right)\nonumber \\
&= \ln \left( \frac{\Phi^2}{\Lambda^2}\right)
\end{align}
up to an additive constant that can be absorbed into $\Phi^2/\Lambda^2$.  Using eqs. (9, 18, 19), we can reduce eq. (14) to 
\begin{equation}
V = \Phi^4 A_0 \left( \lambda \left( \ln  \frac{\Phi^2}{\Lambda^2}\right)\right)
\exp 2 \left[ \int_0^{\lambda\left(\ln \frac{\Phi^2}{\Lambda^2}\right)} dx 
\frac{\gamma(x)}{\beta(x)} + \int_0^\infty dx \frac{fx}{bx^2(1+cx)}\right].
\end{equation}
In eq. (20), all explicit dependence on $\mu^2$ has disappeared; the RG summation of eq. (14) has resulted in a cancellation between the explicit dependence of $V$ on $\mu$ (through $L$) and its implicit dependence on $\mu$ (through $\lambda$ and $\phi^2$).

\section{Renormalization Scheme Dependence in the Massless Scalar Model}

Under the finite renormalizations
\begin{subequations}
\begin{align}
\overline{\lambda} &= \lambda\left(1 + x_1 \lambda + x_2\lambda^2 + \ldots\right)  \\
\overline{\phi}^2 &= \phi^2 \left(1 + y_1 \lambda + y_2\lambda^2 + \ldots \right)
\end{align}
\end{subequations}
it follows that in eqs. (4a,b), $b$, $c$, $f$ are unaltered and that the RS can be characterized by $c_n(n \geq 2)$ and $g_n(n \geq 1)$ [27].  Furthermore, it can be shown that [24, 25, 26]
\begin{subequations}
\begin{align}
\frac{d\lambda}{dc_i} = B_i(\lambda , c_k) &= -b \beta (\lambda) \int_0^\lambda dx \frac{x^{i+2}}{\beta^2(x)} \\
& \approx \frac{\lambda^{i+1}}{i-1} \left[ 1 + \left( \frac{(-i+2)c}{i}\right)\lambda + 
\left( \frac{(i^2-3i+2)c^2+(-i^2+3i)c_2}{(i+1)i}\right)\lambda^2 + \ldots \right]\nonumber \\
\frac{d\lambda}{dg_i} &= 0 \\
\frac{1}{\phi^2} \frac{d\phi^2}{dc_i} = \Gamma_i^c(\lambda) &= \frac{\gamma(\lambda)}{\beta(\lambda)} B_i(\lambda) + b \int_0^\lambda dx \frac{x^{i+2}\gamma(x)}{\beta^2(x)}\\
&\approx \frac{f}{b} \lambda^i \left[ \frac{-1}{i(i-1)} + 2 \left( \frac{c}{i(i+1)}- \frac{g_1}{(i+1)(i-1)}\right) \lambda + \cdots \right]\nonumber \\
\frac{1}{\phi^2} \frac{d\phi^2}{dg_i} = \Gamma_i^g(\lambda) &= f \int_0^\lambda dx \frac{x^{i+1}}{\beta(x)}\\
&\approx \frac{f}{b} \lambda^i \left[ \frac{-1}{i} + \left( \frac{c}{i+1} \right) \lambda + \left( \frac{c_2-c^2}{i+2}\right) \lambda^2 + \cdots \right] .\nonumber
\end{align}
\end{subequations}

From eq. (22), it follows that $\Lambda^2$ in eq. (16) and $\Phi^2$ in eq. (18) are RS invariants under the transformations of eq. (21).

We now use eq. (6) to write eq. (20) as
\begin{equation}
V = \Phi^4 \left( \sum_{n=0}^\infty T_n \lambda^{n+1} \right) \exp 2 \left[ \int_0^\lambda dx \frac{\gamma(x)}{\beta(x)} + \int_0^\infty dx \frac{fx}{bx^2(1+cx)}\right]
\end{equation}
where $T_{n,0} \equiv T_n$ and $\lambda = \lambda\left( \ln \frac{\Phi^2}{\Lambda^2}\right)$.  As $V$ is RS independent, we then have
\begin{align}
\frac{dV}{dc_i} &= \left( \frac{\partial}{\partial c_i} + B_i (\lambda) \frac{\partial}{\partial \lambda} \right) V = 0 \nonumber \\
&= \Phi^4 \exp 2 \Bigg[ \int_0^\lambda dx \frac{\gamma (x)}{\beta(x)} + \int_0^\infty dx \frac{fx}{bx^2(1+cx)}\Bigg] \sum_{n=0}^\infty \Bigg[ \frac{\partial T_n}{\partial c_i} \lambda^{n+1}\nonumber \\
&+ 2b \int_0^\lambda dx \frac{x^{i+2}\gamma(x)}{\beta^2(x)} T_n \lambda^{n+1} + T_n B_i (\lambda) \left( \frac{2\gamma(\lambda)}{\beta(\lambda)} \lambda^{n+1} + (n+1) \lambda^n\right)\Bigg].
\end{align}
Upon using the expansions of eq. (22a,c), eq. (24) leads to
\begin{subequations}
\begin{align}
\frac{\partial T_0}{\partial c_i} &= 0 \\
\frac{\partial T_1}{\partial c_i} &= 0 \\
\frac{\partial T_2}{\partial c_i} &+ \left(-\frac{f}{b} + 1\right) T_0 \delta_2^i  = 0
\end{align}
\end{subequations}
etc. \\
Similarly, we find that
\begin{align}
\frac{dV}{dg_i} &= \frac{\partial V}{\partial g_i} = 0\nonumber \\
&= \Phi^4 \exp 2 \Bigg[ \int_0^\lambda dx \frac{\gamma(x)}{\beta(x)} + \int_0^\infty d\lambda \frac{fx}{bx^2(1+cx)}\Bigg] \sum_{n=0}^\infty \Bigg[ \frac{\partial T_n}{\partial g_i} \nonumber \\
& \qquad + 2 \int_0^\lambda dx \frac{fx^{i+1}}{\beta(x)}T_n \Bigg] \lambda^{n+1}
\end{align}
from which follows
\begin{subequations}
\begin{align}
& \frac{\partial T_0}{\partial g_i}  = 0 \\
& \frac{\partial T_1}{\partial g_i} -\frac{2f}{b} T_0 \delta_1^i  = 0 \\
& \frac{\partial T_2}{\partial g_i} - \frac{f}{b} \left[  T_0 \delta_2^i 
+ \left(2T_1 - cT_0\right) \delta_1^i\right]   = 0 
\end{align}
\end{subequations}
etc.\\
If we integrate eqs. (25, 27) we find that
\begin{subequations}
\begin{align}
T_0 &= \tau_0 \\
T_1 &= \tau_1 + \frac{2f}{b} \tau_0g_1\\
T_2 &= \tau_2 + \left( \frac{f}{b} -1\right) c_2 + \frac{f}{b} \left[ \tau_0 g_2 + \left(2\tau_1 - c\tau_0\right) g_1 + \frac{2f}{b}\tau_0 g_1^2 \right]
\end{align}
\end{subequations}
etc.\\
In eq. (28), $\tau_n$ is a constant of integration and hence is a RS invariant, found by computing $T_0 \ldots T_n$, $g_1 \ldots g_n$, $c_2 \ldots c_n$ in some RS and then solving eq. (28) for $\tau_0 \ldots \tau_n$.  One could now choose $c_i$, $g_i$ so that either $g_i = c_i = 0$, or alternatively, so that $T_n = 0 \, (n \geq 1)$.

\section{Using the Coleman-Weinberg Renormalization Scheme}

When computing $V$, it is often convenient to use the CW RS [6].  For the model of eq. (1), this means that the renormalized coupling $\lambda$ is defined by 
\begin{equation}
\lambda = \left( \frac{d^4V}{d\phi^4}\right)_{\phi = {\mu_{\rm{CW}}}} .
\end{equation}
This condition cannot be satisfied by starting with the $\overline{MS}$ RS and making the transformation of eq. (21).

In the CW scheme
\begin{equation}
V\left(\lambda, \phi^2, \mu^2_{CW}\right) = \sum_{n=0}^\infty \sum_{m=0}^n \overline{T}_{n,m}\, \lambda^{n+1} \overline{L}^m \,\phi^4
\end{equation}
where now
\begin{equation}
\overline{L} = \ln \left( \frac{\phi^2}{\mu^2_{\rm{CW}}}\right)
\end{equation}
in place of eq. (2).  It was noted in ref. [28] that together eqs. (2, 31) provide the relation
\begin{equation}
\mu_{\rm{CW}}^2 = \mu^2/\lambda
\end{equation}
and hence if
\begin{subequations}
\begin{align}
\beta_{\rm{CW}}(\lambda) &= \mu_{\rm{CW}}^2 \frac{d\lambda}{d\mu_{\rm{CW}}^2}\\
\gamma_{\rm{CW}}(\lambda) &= \frac{\mu_{\rm{CW}}^2}{\phi^2} \frac{d\phi^2}{d\mu_{\rm{CW}}^2}
\end{align}
\end{subequations}
we find that [28]
\begin{subequations}
\begin{align}
\beta_{\rm{CW}} &= \frac{\beta}{1-\beta/\lambda}\\
\gamma_{\rm{CW}} &= \frac{\gamma}{1-\beta/\lambda}
\end{align}
\end{subequations}
where in the CW RS $V$ now satisfies the RG equation
\begin{equation}
\left( \mu^2_{CW} \frac{\partial}{\partial\mu^2_{CW}} + \beta_{\rm{CW}} (\lambda) 
\frac{\partial}{\partial\lambda} + \phi^2 \gamma_{\rm{CW}} (\lambda) 
\frac{\partial}{\partial\phi^2}\right) V = 0.
\end{equation}
Together, eqs. (29, 35) can be used to express $V$ entirely in terms of $\beta_{\rm{CW}}$ and $\gamma_{\rm{CW}}$ [29].

Much as with eq. (5) we can make the expansion
\begin{equation}
V = \sum_{n=0}^\infty \overline{A}_n\,(\lambda) \overline{L}^n \,\phi^4 ;
\end{equation}
using the functions
\begin{subequations}
\begin{align}
\hat{\beta}_{\rm{CW}} &= \frac{\beta_{\rm{CW}}}{1-\gamma_{\rm{CW}}}\\
\hat{\gamma}_{\rm{CW}} &= \frac{\gamma_{\rm{CW}}}{1-\gamma_{\rm{CW}}}
\end{align}
\end{subequations}
we can show that
\begin{equation}
V = \overline{\Phi}^4\, \overline{A}_0 \,(\lambda) \exp 2 \left[ \int_0^\lambda dx
\frac{\gamma(x)}{\beta(x)}+ \int_0^\infty \frac{dx fx}{bx^2(1+cx)}\right]
\end{equation}
where $\lambda = \lambda\ln \left( \frac{\overline{\Phi}^2}{\overline{\Lambda}^2}\right)$.  In analogy with eqs. (16, 18) we have
\begin{subequations}
\begin{align}
\ln \left( \frac{\mu^2_{\rm{CW}}}{\overline{\Lambda}^2}\right) &= \int_0^\lambda dx \frac{1}{\beta_{\rm{CW}}(x)} + \int_0^\infty dx \frac{1}{b_{\rm{CW}}x^2 \left( 1 + c_{\rm{CW}}x\right)}\\
\phi^2 &= \overline{\Phi}^2 \exp \left[ \int_0^\lambda dx \frac{\gamma(x)}{\beta(x)} + \int_0^\infty dx \frac{f_{CW}x}{b_{CW}x^2(1+c_{CW}x)}\right]
\end{align}
\end{subequations}
with $\lambda = \lambda\left(\ln \mu^2_{\rm{CW}}/\overline{\Lambda}^2\right)$.  (Recall that by eq. (34), $\gamma_{\rm{CW}}/\beta_{\rm{CW}} = \gamma/\beta$.)  In eq. (38), as in eq. (20), all dependence on the unphysical renormalization scale parameter $\mu^2_{\rm{CW}}$ has cancelled.

\section{Extremizing $V$}

Having found an expression for $V$ in the $\overline{MS}$ and CW RS that depends only on the log-independent contributions to $V$ and is independent of the unphysical renormalization scale parameter, we now will impose the condition that $V(\Phi)$ has an extremum.  From eq. (20), it follows that in the $\overline{MS}$ scheme
\begin{align}
\Phi^2 \frac{dV}{d\Phi^2} &= \Phi^4 \left[ 2\left[1 + \gamma(\lambda)\right] A_0 (\lambda) + \beta(\lambda) A_0^\prime (\lambda)\right] \\
&\qquad \exp 2\left[ \int_0^\lambda dx \frac{\gamma (x)}{\beta(x)} + \int_0^\infty dx \frac{fx}{bx^2(1+cx)}\right]\nonumber
\end{align}
where $\lambda = \lambda\left( \ln \frac{\Phi^2}{\Lambda^2}\right)$.  If this were to vanish at $\Phi = \Phi_0$, then it follows that either
\begin{equation}
\Phi_0 = 0
\end{equation}
in which case there is no spontaneous symmetry breaking, or
\begin{equation}
A_0^\prime (\lambda) + 2 \left( \frac{1+\gamma(x\lambda)}{\beta(\lambda)} \right) A_0 (\lambda) = 0
\end{equation}
when $\lambda = \lambda\left( \ln \frac{\Phi^2_0}{\Lambda^2}\right)$.  This value of $\lambda$ is not fixed, and so $A_0(\lambda)$ must satisfy the differential equation of eq. (42) which leads to
\begin{equation}
A_0(\lambda) = \exp - 2 \left[ \int_0^\lambda dx \frac{1+\gamma(x)}{\beta(x)} + 
\int_0^\infty dx \frac{1+fx}{bx^2(1+cx)}+ K\right]
\end{equation}
where $K$ is a suitably chosen constant of integration.  Together, eqs. (20, 43, 16) result in
\begin{align}
V &= \Phi^4 \exp \left[ -2 \ln \left( \frac{\Phi^2}{\Lambda^2}\right)\right] e^{-2K}\nonumber \\
&= \Lambda^4 e^{-2K}.
\end{align}
We note that if in eq. (43) we use the RS in which $c_i = 0 (i \geq 2)$, $g_i = 0 (i \geq 1)$, then $A_0(\lambda) = \exp\left(\frac{-2}{b\lambda}\right)\left[ \left( \frac{c\lambda}{1+c\lambda}\right)^{\frac{2(f-c)}{b}}\right]$, indicating that there is a non-perturbative contribution to $V$.  Similarly, in the CW RS, it follows from eq. (38) that
\begin{equation}
V = \overline{\Lambda}^4 e^{-2\overline{K}}.
\end{equation}
We find that $V$, if it is to have a non-vanishing extremum, is independent of 
$\phi$-it is ``flat''.  This is consistent with the requirement that $V$ be convex [19-21] and with previously derived results [15-18].

Together, eq. (45) and (29) show that in the CW scheme, $V$ being flat means that $\lambda$ vanishes; it is a ``trivial'' theory [30].  However, the expectation value of $\Phi$ is non-zero; but this expectation value cannot be obtained by locating a local minimum of $V$.  This expectation value can be responsible for mass generation of vector and spinor fields if $\phi$ were to couple to them.  It is also possible that non-trivial contributions to the effective action involving the gradient of $\phi$ can be radiatively generated.

\section{Coupling $\phi$ to a vector field}

To illustrate how the above discussion can be extended to models in which a scalar field is coupled to other fields, we will examine scalar electrodynamics with the action
\begin{equation}
\mathcal{L} = \left( \partial_\mu + ieA_\mu\right) \phi^* \left( \partial^\mu + ieA_\mu\right) \phi - \frac{1}{4} \left( \partial_\mu A_\nu - \partial_\nu A_\mu\right)^2 - \lambda (\phi^*\phi)^2.
\end{equation}
Again, we will only consider the conformal limit in which there is no bare mass term for the scalar field in the classical Lagrangian.  The non-conformal case has been considered in [17, 18].

There are now two couplings, $\lambda$ and $\alpha = e^2/4\pi$, and these have dependence on the RG scale parameter $\mu$ when working in the $\overline{MS}$ RS,
\begin{subequations}
\begin{align}
\mu^2 \frac{d\lambda}{d\mu^2} &= \beta_\lambda (\lambda ,\alpha)\\
\mu^2 \frac{d\alpha}{d\mu^2} &= \beta_\alpha (\lambda ,\alpha)
\end{align}
\end{subequations}
with $\phi^2$ satisfying
\begin{equation}
\mu^2 \frac{d\phi^2}{d\mu^2} = \phi^2 \gamma (\lambda ,\alpha).
\end{equation}
As $V$ is independent of $\mu$, we have
\begin{equation}
\left( \mu^2 \frac{\partial}{\partial \mu^2} + \beta_\lambda \frac{\partial}{\partial\lambda} + \beta_\alpha \frac{\partial}{\partial\alpha} + \phi^2 \gamma \frac{\partial}{\partial\phi^2}\right) V\left( \lambda, \alpha, \phi^2, \mu^2 \right) = 0.
\end{equation}
When using $\overline{MS}$, two types of logarithmic corrections to $V$ can arise, one being $\ln\left(\frac{\lambda\phi^2}{\mu^2}\right)$ and the other 
$\ln\left(\frac{\alpha\phi^2}{\mu^2}\right)$, where $\phi$ is taken to be real.

Following [31] we can write
\begin{equation}
\ln\left(\frac{\alpha\phi^2}{\mu^2}\right) = \ln \left( \frac{\alpha}{\lambda}\right) + \ln\left(\frac{\lambda\phi^2}{\mu^2}\right)
\end{equation}
so that $V$ has explicit dependence on $\mu^2$ only through 
$L =\ln\left(\frac{\lambda\phi^2}{\mu^2}\right)$ .  We then can write
\begin{equation}
V = \sum_{n=0}^\infty A_n(\lambda,\alpha) L^n\phi^4,
\end{equation}
much like eq. (5).  (Other ways of treating these two logarithms appear in refs. [18, 29].)

We now can treat the dependence of $\alpha$ on $\mu$ as being an implicit dependence through its dependence on $\lambda(\mu)$.  From eqs. (47a,b) we know that $\alpha(\lambda)$ is determined by 
\begin{equation}
\frac{d\alpha(\lambda(\mu))}{d\lambda(\mu)} = 
\frac{\beta_\alpha(\lambda(\mu), \alpha(\lambda(\mu)))}{\beta_\lambda(\lambda(\mu), \alpha(\lambda(\mu)))}
\end{equation}
so that
\begin{align}
\beta_\lambda & (\lambda(\mu), \alpha(\mu)) \frac{d}{d\lambda(\mu)} F(\lambda(\mu), \alpha(\lambda(\mu)))\nonumber \\
&= \beta_\lambda  (\lambda(\mu), \alpha(\mu)) 
\left( \frac{\partial}{\partial\lambda(\mu)} + 
\frac{d\alpha(\lambda(\mu))}{d\lambda(\mu)} \frac{\partial}{\partial\alpha(\lambda(\mu))}\right)  F(\lambda(\mu), \alpha(\lambda(\mu)))\nonumber \\
&= \left( \beta_\lambda  (\lambda(\mu), \alpha(\mu)) 
\frac{\partial}{\partial\lambda(\mu)}  + \beta_\alpha  (\lambda(\mu), \alpha(\mu)) 
\frac{\partial}{\partial\alpha(\mu)}\right)F(\lambda(\mu), \alpha(\mu)) .
\end{align}
The RG equation of eq. (47) hence can be written
\begin{align}
& \left( \mu^2\frac{\partial}{\partial\mu^2}  + \beta_\lambda (\lambda(\mu), \alpha(\lambda(\mu))) \frac{d}{d\lambda(\mu)} + \phi^2(\mu) \gamma (\lambda(\mu), \alpha(\lambda(\mu))) \frac{\partial}{\partial\phi^2(\mu)}\right)\nonumber \\
& \hspace{2cm}V\left(\lambda(\mu), \alpha(\lambda(\mu)), \phi^2(\mu), \mu^2\right) = 0
\end{align}
with the function $\alpha(\lambda(\mu))$ satisfying eq. (52).

The deviation of eq. (20) can be generalized so that we find 
\begin{equation}
V = \Phi^4 A_0(\lambda,\alpha(\lambda)) \exp 2 \left[ \int_{K{_\Phi}}^\lambda
dx \frac{\gamma(x, \alpha(x))}{\beta_\lambda(x, \alpha(x))}\right]
\end{equation}
with $\lambda = \lambda\left( \ln \frac{\Phi^2}{\Lambda^2}\right)$
where now
\begin{subequations}
\begin{align}
\ln \left( \frac{\mu^2}{\Lambda^2}\right) &= \int_{K{_\Lambda}}^{\lambda\left(\ln\frac{\mu^2}{\Lambda^2}\right)}
\frac{dx}{\beta_\lambda(x,\alpha(x))}\\
\intertext{and}\nonumber \\
\phi^2& = \Phi^2\exp  \int_{K{_\Phi}}^{\lambda\left(\ln\frac{\mu^2}{\Lambda^2}\right)}
dx \frac{\gamma(x, \alpha(x))}{\beta_\lambda(x,\alpha(x))}.
\end{align}
\end{subequations}

In eqs. (55, 56) $K_\Lambda$ and $K_\Phi$ are positive constants that ensure convergence of the integrals in which they appear.  A shift in their value can be absorbed into $\Lambda$ and $\Phi$.  Integrals involving $f$, $b$ $c$ were used to this end in eqs. (16, 18, 20) but when there is more than one coupling, only the one loop contributions to the RG functions are RS invariant when using a mass independent RS [32, 33] and so this approach cannot be generalized and we employ $K_\Lambda$ and $K_\Phi$.

It is easy to see now from eq. (55) that $V$ has an extremum if $\Phi = \Phi_0$ if either
\begin{equation}
\Phi_0 = 0
\end{equation}
as in eq. (41), or 
\begin{equation}
\frac{d}{d\lambda} A_0(\lambda, \alpha(\lambda)) + 2 \left( \frac{1 + \gamma(\lambda, \alpha(\lambda))}{\beta_\lambda(\lambda,\alpha(\lambda))} \right) A_0(\lambda, \alpha(\lambda)) = 0
\end{equation}
where $\lambda = \lambda\left(\ln \frac{\Phi^2}{\Lambda^2}\right)$.  It then again follows that $V$ must be proportional to the scale parameter $\Lambda^2$ as in eq. (33).

\section{Discussion}

In this paper we have discussed the use of RG summation of logarithmic contributions to the effective potential $V$ using RG summation and shown how this leads to cancellation between the implicit and explicit dependence on the renormalization scale parameter $\mu$. This was done in a conformal model with just a quartic scalar coupling, and conformal scalar electrodynamics.  In the former model, RS dependence within mass independent RS's and the CW RS were also considered.  In both models it was shown that $V$ has an extremum if either there is no spontaneous symmetry breaking or it is constant.  This result follows from examining the RG summed form of $V$.  The cosmological consequences are worth examining [34].

We would like to extend the approach used here to models which are non-conformal, involve the Stueckelberg mechanism [1, 2] or in which there are several scalar fields\vspace{1cm}.

\noindent
{\Large\bf{Acknowledgements}}\\
The author would like to thank F.T. Brandt, F.A. Chishtie and T.N. Sherry for many conversations on the renormalization group and its application to the effective potential.  Also, Roger Macleod made several useful comments.

\end{document}